\documentclass[12pt]{article}
\usepackage{cite}
\begin{document}


\title{\large\bf Quantum spin ladder systems associated with $su(2|2)$. }


\author{A. Foerster$^1$\thanks{angela@if.ufrgs.br},
K.E. Hibberd$^2$\thanks{keh@cbpf.br},
J.R. Links$^3$\thanks{jrl@maths.uq.edu.au} and
I. Roditi$^4$\thanks{roditi@cbpf.br}.}

\maketitle        

\begin{center}

${}^1$Instituto de F\'{\i}sica da UFRGS,\\
Av. Bento Gon\c{c}alves 9500, Porto Alegre, RS - Brazil.\\
\vspace{.25cm}
${}^{2,4}$Centro Brasileiro de Pesquisas F\'{\i}sicas, \\
Rua Dr. Xavier Sigaud 150, 22290-180, Rio de Janeiro, RJ - Brazil.\\
\vspace{.25cm}
${}^3$ Centre for Mathematical Physics,\\
Department of Mathematics,\\
The University of Queensland, 4072, Australia. \\

\vspace{.5cm}

\end{center}
\maketitle

\vspace{10pt}

\begin{abstract}

Two integrable quantum spin ladder systems will be introduced associated with the fundamental $su(2|2)$ solution of the Yang-Baxter equation.  The first model is a generalized quantum Ising system with Ising rung interactions.  In the second model the addition of extra interactions allows us to impose Heisenberg rung interactions without violating integrability.  The existence of a Bethe ansatz solution for both models allows us to investigate the elementary excitations for antiferromagnetic rung coupling
s.  We find that the first model does not
show a gap whilst in the second case there is a gap for all positive values of the rung coupling.

\end{abstract}



\clearpage


\def\a{\alpha}
\def\b{\beta}
\def\d{\delta}
\def\e{\epsilon}
\def\g{\gamma}
\def\k{\kappa}
\def\l{\lambda}
\def\o{\omega}
\def\t{\tau}
\def\s{\sigma}
\def\D{\Delta}    
\def\L{\Lambda}


\def\beq{\begin{equation}}
\def\eeq{\end{equation}}
\def\bea{\begin{eqnarray}}
\def\eea{\end{eqnarray}}
\def\ba{\begin{array}}
\def\ea{\end{array}}
\def\no{\nonumber}
\def\le{\langle}
\def\re{\rangle}
\def\lt{\left}
\def\rt{\right}
\def\dwn{\downarrow}   
\def\up{\uparrow}
\def\dag{\dagger}
\def\nonum{\nonumber}

\newcommand{\reff}[1]{eq.~(\ref{#1})}

\vskip.3in


The study of spin ladder systems continues to 
receive a great deal of interest from both theoretical and experimental 
points of view providing new insights into low dimensional quantum systems.  
By now it is well established that the
Heisenberg spin-1/2 ladders with an even number of legs are found to have 
finite energy gap while those with an odd number of legs
show gapless spin excitations and these results have been 
verified experimentally in a number of systems \cite{dr}.
Under hole-doping these ladders are predicted to pair and may superconduct.
The quantum phase transition between the gap and gapless phase also proves to be
both theoretically and experimentally interesting.
Several generalized ladders incorporating interactions beyond simple rung and leg exchange 
have appeared in the literature and demonstrate remarkably interesting behaviour 
in relation to the ground state structure and excitation spectra (see for example \cite{nt,km}).

Another approach has been to derive quasi two-dimensional systems using the
well established theories from one-dimension.
A number of integrable ladder systems involving tunable interaction parameters
can been found in the literature  for which 
some thermodynamic quantities have been obtained \cite{mg,mt,fr1,pz,afw,fk,szh,bm1,bm2,gbm,flt,lf}.
An example is the two-leg zigzag  Majumdar-Ghosh ladder \cite{mg}, which
was generalized by \cite{mt} 
using higher order conservation laws of the Heisenberg chain to obtain 
an exactly solvable model.  Along the same lines, Frahm and R\"odenbeck used 
an algebra homomorphism to develop
an integrable family of coupled Heisenberg spin-1/2 chains \cite{fr1}.

Various ladder models have been developed by
an extension of the symmetry algebra \cite{afw,fk,bm1,bm2,gbm,w}.
Families of n-leg spin-1/2 ladders were obtained from $su(N)$ spin chains in 
\cite{bm1} and this work was extended in \cite{bglm}.
An $su(2)$-invariant spin-1/2 ladder model was introduced in  
\cite{afw} and showed both isotropic 
exchange and biquadratic interactions.  A solution of 
this model was obtained in \cite{bm2} by showing 
that the model can be made to satisfy the Hecke algebra.
Another example is the ladder model 
formulated using the Hubbard model in \cite{lf}.  
A new set of novel spin ladder models possessing extra free parameters 
were introduced by \cite{flt} which were shown to reduce to those of Wang \cite{w} for the $su(4)$ and $su(3|1)$.

In this work we present two new solvable ladder systems with 
coupled rung 
interactions.  We begin with the $su(2|2)$ invariant solution of the Yang-Baxter equation \cite{EKS} which under a Jordan-Wigner transformation can be considered as a generalized quantized Ising model.
Ising rung interactions can be coupled into the model without corrupting the integrability of the system.  However, an attempt to directly introduce Heisenberg interactions fails.  To overcome this, we apply a basis transformation, which yields a Hamilton
ian which does
commutes with the Heisenberg rung interaction.
We present the Bethe ansatz solution and the resulting expression for the energy is obtained.  An analysis of the Bethe ansatz solution shows that for the generalized Ising case there is no gap for the elementary excitations.  On the other hand, the model
 with Heisenberg interactions does show a gap for all positive values of the coupling parameter $J$.  This situation is in some contrast to other recently proposed generalized spin ladder systems formulated within the quantum inverse scattering method (QISM) \cite{gbm,lf,w}.  In these cases the gap closes at some critical value greater than zero.  Numerical and experimental results suggest that the gap should prevail for all positive values of $J$ \cite{dr}.

The first model has a local Hamiltonian of the following form
\bea
H_{i,j}&=&
\left[\s_i^+\s_j^-+\s_i^-\s_j^+ +1/2(\s_i^z+\s_j^z)\right]
\left[\t_i^+\t_j^-+\t_i^-\t_j^+ +1/2(\t_i^z+\t_j^z)\right] \nonum \\
&&+ 
J/2 (\s_i^z \tau_i^z + \s_j^z \t_j^z), 
\label{h1}
\eea
where $J$ above can take arbitrary real values.  The sets $\{ \s^{\pm}_j, \s_j^z \}$ and $\{ \t^{\pm}_j, \t_j^z \}$ are two commuting sets of Pauli matrices acting on site $j$ which describe interactions on the two legs of the ladder.

The global Hamiltonian 
$$H=\sum_{j=1}^N H_{j,j+1}$$
with periodic boundary conditions is integrable as a result of the QISM.  In this case, the $R$-matrix satisfying the Yang-Baxter equation is the fundamental $su(2|2)$ solution \cite{EKS} from which the Hamiltonian is derived in the usual manner through the QISM for $J=0$.  It is pertinent to mention at this point that the $R$-matrix solution of the Yang-Baxter equation is used in an ungraded context here.  This is achieved by means of the well known Jordan-Wigner transformation mapping between canonical 
spin and fermi operators.
We can arbitrarily couple (with coupling $J$) the Ising rung interactions in the model above by virtue of the fact that these terms commute with the Hamiltonian (\ref{h1}).  

In order to obtain a model with Heisenberg rung interactions we introduce additional terms into the Hamiltonian which now reads
\bea
\check{H}& =&\sum_{j=1}^N\left\{\left[\s_j^+\s^-_{j+1}+\s_j^-\s_{j+1}^+ +1/2(\s_j^z+\s_{j+1}^z)\right]
\left[\t_j^+\t_{j+1}^-+\t_j^-\t_{j+1}^+ +1/2(\t_j^z+\t_{j+1}^z)\right]\right.\nonum\\
&& -  
( \s_{j}^+ \t_{j}^- + \s_{j}^- \t_{j}^+ ) 
( \s_{j+1}^+ \t_{j+1}^- + \s_{j+1}^- \t_{j+1}^+ )  -
 1/4 (\s_j^z -\t_j^z) (\t_{j+1}^z - \s_{j+1}^z) \nonum \\
&&\left.+ J/2 (\vec{\s_j} . \vec{\t_j}+\vec{\s_{j+1}} . \vec{\t_{j+1}})
 \right\}.
\label{transformedh}
\eea
In the absence of rung interactions, the local Hamiltonians for the two models are related through the following basis transformation
\bea
 |\up,\up\re&\rightarrow&  |\up,\up\re, \nonum \\
|\up,\dwn\re &\rightarrow&1/\sqrt{2} ( |\up,\dwn\re - |\dwn,\up\re), \nonum \\
|\dwn,\up \re &\rightarrow& 1/\sqrt{2} ( |\up,\dwn\re + |\dwn,\up\re), \nonum \\
|\dwn,\dwn\re &\rightarrow&|\dwn,\dwn\re.
\label{basistransf}
\eea
Note that it is possible to include more general interactions in both models, such as an applied magnetic field, which will commute with these Hamiltonians.  However, here we will limit ourselves to only those described above.

It can been seen for both models that besides the rung interaction, there are Ising interactions along the legs and in addition there exist $XX$ correlated exchanges and four-body biquadratic $XX$ exchange interactions.  In the absence of rung interactions both models retain nontrivial spin interactions between the legs.  

From the Bethe ansatz solution which we present below, we can determine  
the behaviour of the elementary excitations for the systems.
For the Hamiltonian (\ref{h1}), the energy levels are given by 
\bea
E  =  \sum^{n_1+n_2+n_3}_{j=1} \frac 1{\l_j^2+1/4}-N-  J(N - 2 n_2-2n_3), 
\label{energy1}
\eea
while for the second model (\ref{transformedh}) they read
\bea
\check{E}  =  \sum^{n_1+n_2+n_3}_{j=1} \frac 1{\l_j^2+1/4}
-N-J[3 N  -4( n_1+n_2+n_3)]. 
\label{energytrf}
\eea
Above $N$ is the lattice length and $n_1, ~n_2,~n_3$ are quantum numbers for the eigenstates.  In both cases the variables
$\l_j$ are solutions of the following Bethe ansatz equations (BAE). 
These equations are different from those given for example in \cite{bae}
since the Jordan-Wigner transformation from fermi operators to spin 
operators induces sector dependent twisted boundary conditions. However, 
the procedure to obtain these equations in this case follows in 
much the same way so we simply give the final result
\bea
\left( \frac{\l_k - i/2}{\l_k + i/2 } \right)^N &=&
 \prod^{n_2+n_3}_{j=1} \frac{ \l_k-\mu_j+i/2}{  \l_k-\mu_j-i/2}
\prod^{n_1+n_2+n_3}_{l=1,l\neq k} \frac{ \l_l-\l_k+i}{  \l_l-\l_k-i},
~~~k=1,..., n_1+n_2+n_3,\nonum \\
\prod^{n_3}_{m=1} 
\frac{ \nu_m-\mu_j+i/2}{  \nu_m-\mu_j-i/2}&=&
-(-1)^{(N+n_2+n_3)}\prod^{n_1+n_2+n_3}_{k=1}
\frac{ \l_k-\mu_j+i/2}{  \l_k-\mu_j-i/2}
,~~~j=1,...,
n_2+n_3,\nonum \\
\prod^{n_3}_{l=1,l\neq m} \frac{ \nu_l-\nu_m+i}{  \nu_l-\nu_m-i} &=& 
\prod^{n_2+n_3}_{j=1} \frac{ \nu_m-\mu_j-i/2}{  \nu_m-\mu_j+i/2},
~~~m=1,...,n_3. 
\label{BAE}
\eea

For the first model (\ref{h1}), there is no gap when $J=0$.  For large values of $J$, in which case the rung interactions dominate, there is still no gap due to the non-uniqueness of the ground state for a two-site Ising system.  This observation is confirmed from the BAE.  An elementary excitation is characterized by the quantum numbers $n_2=n_3=0$ and $n_1=1$.  Solving the BAE yields the energy expression from (\ref{energy1})
\bea
E  =   \frac 1{\l^2+1/4}-N-  JN, 
\label{energyex}
\eea
where $\lambda = 1/2 \cot (\pi k/N),~ k=1,...,N-1$.  Choosing $k=1$, then the thermodynamic limit $N \rightarrow \infty $ shows that  $\l \rightarrow \infty$ and hence there is no gap.

Now we turn our attention to the second model (\ref{transformedh}).  
Again there is no gap for $J=0$.  In the limit of large $J$, the ground 
state consists of 
a product of rung singlets indicating a gap to any excited state.  
Using the same solution of the BAE as in the above case, the energy of an excited state in this model is given from (\ref{energytrf}) by 
\bea
\check{E } = \frac 1{\l^2+1/4}-N-J(3 N -4). 
\label{energytrfex}
\eea
It is apparent that there is a gap of 
$$ \Delta =  \frac 1{\l^2+1/4} +4 J. $$
In the thermodynamic limit we can allow $\l \rightarrow \infty $ which shows that a gap of $\Delta= 4J$ persists.
Our analysis shows that there is always a gap for this model in the case of antiferromagnetic Heisenberg interactions along the rungs of the ladder.  Only in the limit of vanishing coupling $J$ does the gap close.

\vspace{.5cm}

{\bf Acknowledgements:}  We thank the Australian Research Council,
FAPERJ (Funda\c{c}\~{a}o de Amparo a Pesquisa do Estado do Rio de Janeiro) and CNPq (Conselho Nacional de Desenvolvimento Cient\'{\i}fico e Tecnol\'ogico) for financial support. 
KEH would like to thank the Centre for Mathematical Physics at the University of Queensland for their generous hospitality.

\end{document}